\documentclass[twocolumn,showpacs,preprintnumbers,amsmath,amssymb,pra,floatfix]{revtex4} 

\usepackage{graphicx}
\usepackage{dcolumn}
\usepackage{longtable}
\usepackage{bm}
\usepackage[mediumqspace,mediumspace,amssymb,Gray]{SIunits}
\usepackage{color}
\begin{document}
\preprint{  }
%
%

\title{Production and decay of Sulphur excited species in a ECRIS plasma}


%
%
%
\author{M.\ C.\ Martins} \email{mdmartins@fc.ul.pt} \author{J.\ P.\ 
  Marques} \email{jmmarques@fc.ul.pt} \author{A.\ M.\ Costa}
\email{amcosta@fc.ul.pt}
\affiliation{Centro de F{\'\i}sica At{\'o}mica, Departamento de F{\'\i}sica, Faculdade de Ci{\^e}ncias, \\
  FCUL, Universidade de Lisboa, Campo Grande, Ed. C8, 1749-016 Lisboa,
  Portugal }
\author{J.\ P.\ Santos} \email{jps@fct.unl.pt} \author{ F.\ Parente}
\email{facp@fct.unl.pt} \affiliation{Centro de F{\'\i}sica
  At{\'o}mica, Departamento de F{\'\i}sica, Faculdade de Ci{\^e}ncias
  e
  Tecnologia, \\
  FCT, Universidade Nova de Lisboa, 2829-516 Caparica, Portugal }
\author{S. Schlesser, E.-O. Le Bigot, P.\ Indelicato}
\email{paul.indelicato@spectro.jussieu.fr}
\affiliation{Laboratoire Kastler Brossel, \'Ecole Normale Sup\' erieure; 
  CNRS; Universit\' e P. et M. Curie - Paris 6, Case 74; 4, place
  Jussieu, 75252 Paris CEDEX 05, France
}
  
\date{\today}
%

\begin{abstract}
  The most important processes for the creation of S$^{12+}$ to
  S$^{14+}$ ions excited states from the ground configurations of
  S$^{9+}$ to S$^{14+}$ ions in an electron cyclotron resonance ion
  source, leading to the emission of K X-ray lines, are studied.
  Theoretical values for inner-shell excitation and ionization cross
  sections, including double KL and triple KLL ionization, transition
  probabilities and energies for the deexcitation processes, are
  calculated in the framework of the multi-configuration Dirac-Fock
  method. With reasonable assumptions about the electron energy
  distribution, a theoretical K$\alpha$ X-ray spectrum is obtained,
  which is compared to recent experimental data.
\end{abstract}

%
%
%
%
\maketitle
%
%
%
%
\section{Introduction}
\label{sec:intro}
Electron-cyclotron resonance ion sources (ECRIS) are characterized by
their capacity to produce large populations of highly charged ions and
by high electron temperatures. X-ray emission, including
bremsstrahlung and characteristic lines, caused by these high-energy
electrons due to electron--ion collisions, has been used for plasma
diagnostics. 

In 2000, Douysset \textit{et al.} \cite{Douysset1} proposed to
estimate the ionic density of each charge state in an ECRIS plasma
through the measurement of the intensity of the emitted K$\alpha$
lines. As pointed out by these authors, this calculation depends,
 on the following conditions:
\begin{itemize}
\item precise identification of the X-ray lines emitted by the
  different charge states;
\item a deep understanding of the different paths leading to the
  excited states and their cross sections;
\item knowledge of the shape of the electronic distribution function
  above the threshold for production of excited states.
\end{itemize}

In 2001, some of us published~\cite{Martins1} an analysis of K X-ray spectra
emitted by Ar ions in an ECRIS plasma~\cite{Douysset1}, showing that a
complete analysis of these spectra calls for a careful examination of
all excitation and ionization processes that lead to the excited
states of the different ionic species whose decay will yield the
detected lines. We showed that single-ionization processes were not
enough to explain all the observed features and only when double KL
ionization is taken into account is it possible to account for all of
them.

Several ECR ion sources with superconducting coils and superconducting
or permanent magnet hexapoles are now becoming available
\cite{lbbg1998,bsh2000,llac2005,cgbc2008}. One of them, the
Electron-Cyclotron resonance ion trap (ECRIT) at the Paul Scherrer
Institute \cite{bsh2000}, has been specifically designed for X-ray
spectroscopy of the ions in the plasma. This ECRIT has a large mirror
ratio, a large plasma volume, and is an ideal instrument for detailed
observation of the plasma for highly-charged medium-$Z$ ions.  Even
though the frequency of the injected microwave was low (6.4 GHz),
X-rays of core-excited ions ranging from heliumlike to quasi neutral
sulfur, chlorine and argon have been measured.  Thanks to this source
and a high-resolution spherically-bent crystal X-ray spectrometer,
high-resolution spectra ($\approx$~\unit{0.3}{eV}) are now available
\cite{Indelicato07,Indelicato08}, which represent an order of
magnitude improvement over the previous experiment \cite{Douysset1}
This improved resolution calls for a theoretical study more refined
than the one we used for the previous work, and allow for the
identification of new mechanism, as more lines can be resolved.

In the present work, we examine the more important atomic processes
that may take place in the PSI ECRIT, and contribute to the creation
of low-lying excited states with a K hole which will de-excite by the
emission of K X-ray lines, in order to find out the ionic density of
each charge state in a plasma. We used a multi-configuration
Dirac-Fock (MCDF) program, including intra- and outer-shell
correlation, relativistic and quantum electrodynamics (QED)
contributions, to obtain values of energies and transition
probabilities for the subsequent de-excitation processes.  Theoretical
values for excitation and ionization cross sections, including double
and triple ionization in K- and L-shells, were computed.

Combining these results with estimated values of ion densities, we
were able to generate theoretical K$\alpha$ X-ray spectra.  A
comparison with an experimental spectrum has led to more realistic
values of ion densities.

This paper is organized as follows. In Sec. \ref{sec:mcdf}, we present
briefly the MCDF method, with emphasis on the specific features of the
code which are important for energy correlation calculations. In Sec.
\ref{sec:analysis}, we analyse sulphur spectra obtained in an ECRIS
detailing the experiment, the peak energies calculations, the
processes leading to the creation of ionic excited states, and the
calculation of the corresponding cross sections, and finally the
calculation of the line intensities. In Section \ref{sec:Res} we
present the results and discuss them.

%
%
%
\section{The MCDF method}
\label{sec:mcdf}
The general relativistic MCDF code developed by J. P. Desclaux and P.
Indelicato~\cite{descl01,Desclaux75,mcdf} was used to calculate
bound-state wave functions and energies. Details of the method,
including the Hamiltonian and the processes used to build the wave
functions can be found elsewhere~\cite{Grant88,Indelicato95}.

The total wave function is calculated with the help of the variational
principle. The total energy of the atomic system is the eigenvalue of
the equation
\begin{equation}
\label{eq001}
        {\cal H}^{\mbox{{\tiny no pair}}}
        \Psi_{\Pi,J,M}(\ldots,\mathbf{r}_{i},\ldots)=E_{\Pi,J,M}
        \Psi_{\Pi,J,M}(\ldots,\mathbf{r}_{i},\ldots),
\end{equation}
where $\Pi$ is the parity, $J$ is the total angular momentum
eigenvalue, and $M$ is the eigenvalue of its projection on the $z$
axis $J_{z}$. In this equation, the hamiltonian is given by
\begin{equation}
{\cal H}^{\mbox{{\tiny no pair}}} = \sum_{i=1}^{N} {\cal H}_D (r_i) + \sum_{i<j} V_{ij} (|\mathbf{r}_{ij}|),
\end{equation}
where ${\cal H}_D$ is the one electron Dirac operator and $V_{ij}$ is
an operator representing the electron-electron interaction of order
one in $\alpha$. The expression of $V_{ij}$ in Coulomb gauge, and in
atomic units, is
\begin{eqnarray}
\label{eq:coulop}
V_{ij} =& \,\,\,\, \frac{1}{r_{ij}}  \\
\label{eq:magop}
&-\frac{\boldsymbol{\alpha}_{i} \cdot \boldsymbol{\alpha}_{j}}{r_{ij}} \\ 
\label{eq:allbreit}
& - \frac{\boldsymbol{\alpha}_{i} \cdot
         \boldsymbol{\alpha}_{j}}{r_{ij}} 
[\cos\left(\frac{\omega_{ij}r_{ij}}{c}\right)-1]
         \nonumber \\
& + c^2(\boldsymbol{\alpha}_{i} \cdot
         \boldsymbol{\nabla}_{i}) (\boldsymbol{\alpha}_{j} \cdot
         \boldsymbol{\nabla}_{j})
         \frac{\cos\left(\frac{\omega_{ij}r_{ij}}{c}\right)-1}{\omega_{ij}^{2} r_{ij}},
\end{eqnarray}
where $r_{ij}=\left|\mathbf{r}_{i}-\mathbf{r}_{j}\right|$ is the
inter-electronic distance, $\omega_{ij}$ is the energy of the
exchanged photon between the two electrons, $\mathbf{\alpha}_{i}$ are
the Dirac matrices and $c$ is the speed of light. We use the Coulomb
gauge as it has been demonstrated that it provides energies free from
spurious contributions at the ladder approximation level and must be
used in many-body atomic structure
calculations~\cite{Gorceix88,Lindroth89}.

The term (\ref{eq:coulop}) represents the Coulomb interaction, the
term (\ref{eq:magop}) is the Gaunt (magnetic) interaction, and the
last two terms (\ref{eq:allbreit}) stand for the retardation operator.
In this expression the $\boldsymbol{\nabla}$ operators act only on
$r_{ij}$ and not on the following wave functions.

The MCDF method is defined by the particular choice of a trial
function to solve the Dirac equation as a linear combination of
configuration state functions (CSF):
\begin{equation}
\left\vert \mathit{\Psi}_{\mathit{\Pi},J,M}\right\rangle =\sum_{\nu=1}%
^{n}c_{\nu}\left\vert \nu,\mathit{\Pi},J,M\right\rangle . \label{eq_cu}%
\end{equation}
The CSF are also eigenfunctions of the parity $\mathit{\Pi}$, the
total angular momentum $J^{2}$ and its projection $J_{z}$. The label
$\nu$ stands for all other numbers (principal quantum number, etc.)
necessary to define unambiguously the CSF. The $c_{\nu}$ are called
the mixing coefficients and are obtained by diagonalization of the
Hamiltonian matrix coming from the minimization of the energy in
Eq.~\eqref{eq001} with respect to the $c_{\nu}$.

The CSF are antisymmetric products of one-electron wave functions
expressed as linear combination of Slater determinants of Dirac
4-spinors
\begin{equation}
\left\vert \nu,\mathit{\Pi},J,M\right\rangle =\sum_{i=1}^{N_{\nu}}%
d_{i}\left\vert
\begin{array}
[c]{ccc}
\psi_{1}^{i}\left(  r_{1}\right)  & \cdots & \psi_{m}^{i}\left(  r_{1}\right)
\\
\vdots & \ddots & \vdots\\
\psi_{1}^{i}\left(  r_{m}\right)  & \cdots & \psi_{m}^{i}\left(  r_{m}\right)
\end{array}
\right\vert .
\end{equation}
where the $\psi$-s are the one-electron wave functions and the
coefficients $d_{i}$ are determined by requiring that the CSF is an
eigenstate of $J^{2}$ and $J_{z}$. The $d_{i}$ coefficients are
obtained by requiring that the CSF are eigenstates of $J^{2}$ and
$J_{z}.$

The Multi-Configuration approach is characterized by the fact that a
small number of configurations can account for a large amount of
correlation.

The so-called Optimized Level (OL) method was used to determine the
wave function and energy for each state involved. In this way,
spin-orbitals in the initial and final states for the radiative
transitions are not orthogonal, since they have been optimized
separately. This non-orthogonality effect is fully taken into account
\cite{indelicato96,indelicato97}, using the formalism proposed by
L\"{o}wdin \cite{lowdin55}. The length gauge has been used for all
radiative transition probabilities.

Radiationless transition probabilities were calculated using
Desclaux's code \cite{Santos1}. The bound wave functions were
generated using this code for configurations that contain one initial
inner-shell vacancy while the continuum wave functions were obtained
by solving the DF equations with the same atomic potential of the
initial state. With this treatment, the continuum wave functions are
made orthogonal to the initial bound state wave functions, thus
assuring orthogonality.  The continuum wave function is normalized to
represent one electron per unit time.

QED contributions for the transition energies and transition
probabilities have been found to be negligible. For example, the QED
contributions for the $1s 2s \to 1s^2$ transition energy and
probability are, respectively, 0.03\% and 0.01\%.
%
%
\section{Analysis of sulphur spectra in an
  ECRIS plasma} 
\label{sec:analysis}
%
%
\subsection{Experiment}
\label{sec:exp}
Sulphur X-ray spectra were obtained in the Electron 6.4 GHz Cyclotron
Resonance Ion Trap (ECRIT) at the Paul Scherrer Institut
(PSI)~\cite{Trassinelli07}, by the Pionic Hydrogen Collaboration
 (see \url{http://www.fz-juelich.de/ikp/exotic-atoms/index.php}). The experimental set-up was composed
mainly of two parts: the ECR ion trap (the X-ray source) \cite{bsh2000} and a Bragg
spectrometer set-up in Johann geometry \cite{gaab1999,Anagnostopoulos05}. The X rays reflected by the
spectrometer crystal were recorded by a two-dimensional
position-sensitive detector \cite{naab2002} placed in the proximity of the Rowland
circle of the spectrometer. 

In order to obtain X-ray spectra from highly-charged sulphur, SO$_{2}$
gas was injected. To improve the ionization efficiency, a gas mixture
with O$_{2}$ was used, adjusted to a mixing ratio of about 1:9 (the main
gas is oxygen). Details about calibration and spectra construction
can be found in Refs.~\cite{Trassinelli05,Trassinelli07,itab2008}.

The obtained spectra cover an energy region corresponding to He-like,
Li-like, and Be-like K$\alpha$ line energies of sulphur ions.

%
%
\subsection{Transition energies}
\label{sec:peak}
The spectra obtained at PSI cover the 2.400--\unit{2.460}{keV} energy
range.  The more important features in these spectra are the Be-like
$1s 2s^2 2p\,^1P_1 \to 1s^22s^2\,^1S_0$ E1 line at \unit{2.418}{keV},
the He-like $1s 2s\,^3S_1 \to 1s^2\,^1S_0$ M1 line at
\unit{2.430}{keV}, and the Li-like $1s 2s 2p \,^2P_{J} \to 1s^2 2s
\,^2S_{1/2}$, $J=1/2, 3/2$ lines, at \unit{2.437}{keV} and
\unit{2.438}{keV}, respectively.

In a preliminary calculation, we have used the MCDF code, including
only intra-shell correlation, to calculate the energies of these lines
in the same way as in our previous calculations for Ar ions
spectra~\cite{Martins1}. The basis space includes all possible
electron configurations built from $1s$, $2s$, and $2p$ orbitals
corresponding to single and double excitations from the main
configuration.
The experimental spectrum energy scale has been fixed by setting the
He-like $1s 2s\,^3S_1 \to1s^2\,^1S_0$ M1 transition to
\unit{2430.351}{eV}, using the high-precision QED calculation from
Ref.~\cite{asyp2005}.  In this case the theoretical results differ
from the experimental ones by \unit{0.46}{eV} and \unit{0.41}{eV} for
the two Li-like lines, respectively, as shown in Tab.~\ref{tab:Corr}
and in Fig.~\ref{fig:monoion}. These discrepancies are due to the
higher accuracy of the present experimental data. Moreover the
improved resolution (\unit{0.3}{eV}) in comparison with the Ar case
(\unit{3}{eV}) makes the comparison between theoretical and
experimental spectra more sensitive to shifts. Therefore a new
calculation was performed, including correlation up to the $4f$
sub-shell. For that purpose, the basis space was enlarged to include
now all orbitals in the $n=3,4$ electronic shells. A much better
agreement with experiment is obtained (see Tab.~\ref{tab:Corr}), as
the energy differences between theory and experiment were reduced to
\unit{0.06}{eV} and \unit{0.04}{eV}, respectively, for the two Li-like
lines, thus showing the importance of electronic correlation in the
interpretation of high resolution experiments.  We are dealing here
with auto-ionizing initial states in these transitions, except in the
case of heliumlike line.  An auto-ionizing state correspond to a
resonance degenerate with a continuum, corresponding to a final state
with a free electron, which lead to both a broadening (due to the
Auger transition probability) and a shift. Up to now this shift has been
 evaluated only for neutral atoms with an inner-shell hole
\cite{ibl1998,dkib2003}. Here, a rough estimate of this effect shows
it must be in the order of a few meV to a few tenth of eV, depending on the
number of allowed Auger channels.

Using our calculation, including only correlation up to $n=4$, we
obtain for the M1 line the energy \unit{2430.28}{eV}, \unit{0.07}{eV}
below the experimental value.  A more accurate energy shift could be
deduced, if the experiment was calibrated against our MCDF value, as
some cancellations in the correlation energy should occur: both the initial and final states
of the main Li-like transitions, for example, correspond to the M1 line initial and final states with an extra electron.

%
%
\subsection{Excited species production}
\label{sec:exc}

Radiative spectra obtained in ECRIS systems usually include radiative
lines resulting from de-excitation of ions in different charge states
and different levels. A detailed analysis of the main processes
leading to those initial states is needed. We refer the reader to
Martins \textit{et al.} \cite{Martins1} for details.

We assume that all S$^{q+}$ ions, where $q$ is the degree of
ionization ($q=Z-m$, $m$ being the number of electrons in the ion),
are initially in the ground configuration. The two main processes
leading to a hole in the K shell are K ionization of the S$^{\left(
    q-1\right) +}$ ion ground configuration and excitation of the
S$^{q+}$ ion ground configuration. In order to generate a preliminary
theoretical spectrum, we took into account processes that create
single K-hole excited states with two, three, and four electrons from
the ground configurations of sulphur ions.

The intensity of the line corresponding to a transition from level $i$
to a level $j$ in a S$^{q+}$ ion with a K hole and in level $i$, is
given by
\begin{equation}
\label{eq002}
I_{ij}^{q} = \hbar\omega A_{ij}^q  N_i^{\textrm{\tiny K},q}.
\end{equation}
In this equation $\hbar\omega$ and $A_{ij}^{q}$ are the radiative
$i\to j$ transition energy and probability, respectively, and
$N_i^{\textrm{\tiny K},q}$ is the density of S$^{q+}$ ions with a K
hole and in level $i$.  This density is obtained from the balance
equation
\begin{equation}
\label{eq003}
N_{0}^{q}\left\langle N_e v_e\sigma_{i}^{\textrm{\tiny K-exc,}q}\right\rangle
+ N_{0}^{q-1}\left\langle N_e v_e\sigma_{i}^{\textrm{\tiny K-ion,}\left(
q-1,q\right)  }\right\rangle  = N_i^{\textrm{\tiny K},q}A_{i}^{q}.
\end{equation}
Here, $N_e$ is the electron density, $N_0^q$ and $N_0^{q-1}$ are the
densities of S$^{q+}$ and S$^{(q-1)+}$ ions, respectively, in the
ground configuration, $A_{i}^{q}$ is the transition probability for
de-excitation of these ions by all possible processes (radiative and
radiationless), $\sigma_{i}^{\mathrm{K-exc},q}$ is the excitation
cross section for the processes leading from the S$^{q+}$ ion in the
ground configuration to the excited level $i$ of the same ion with a K
hole, and $\sigma_{i}^{\mathrm{K-ion},\left( q-1,q\right)}$ is the
ionization cross section leading from the S$^{(q-1)+}$ ion in the
ground configuration to the excited level $i$ of the S$^{q+}$ ion with
a K hole.

The quantities $\langle N_e v_e\sigma_{i}\rangle$ give the rate of the
number of events related to a process (excitation or ionization),
averaged over the electron distribution energy, and are defined by
\begin{equation}
\langle N_e v_e\sigma_{i}\rangle = N_e \int_{E_{min}}^\infty f_e(E)
v_e(E) \sigma (E) dE, 
\end{equation}
where $f_e(E)$ is the electron distribution energy, and $v_e(E)$ and
$\sigma (E)$ are, respectively, the electron velocity, and the cross
section for the process, for electron energy $E$.

Electron-impact excitation cross sections were computed using the
first Born approximation following the work of Kim \textit{et al.}
\cite{Kim258,Kim259}. In these calculations we obtained the cross sections for
the processes leading from each level $j$ of the S$^{q+}$ ion ground
configuration, to the excited level $i$ of the S$^{q+}$ ion with a
K-shell hole, $\sigma_{ji}$. We used a
Multi-Configuration Dirac-Fock (MCDF) wave function for the atom and a
Dirac wave function for the free electron. Only the Coulomb
interaction between the free electron and the atomic electrons was
considered. The individual cross sections thus obtained were then
weighted by the statistical weight $g_j$ of each $j$ level of the
ground configuration in order to obtain the excitation cross section
$\sigma_{i}^{\mathrm{K-exc},q}$.

K-shell ionization cross-sections were computed using the relativistic
binary-encounter-dipole (RBED) model \cite{Kim00}. This method has
provided accurate results \cite{Santos03}, even at energies close to
the ionization threshold.  We use a simplified version of this method,
referred as the relativistic binary-encounter-Bethe (RBEB) model, in
which, besides the incident electron kinetic energy $T$, only three
orbital constants are needed: the kinetic energy of the ejected
electron $U$, the orbital binding energy $B$, and the electron
occupation number $N$ for the pertinent shell. The RBEB cross section
expressions reads
\begin{widetext}
\begin{eqnarray}
 \sigma_{\rm RBEB} &=& \frac {4\pi a_0^2\alpha^4N}
{(\beta_t^2+\beta_u^2+\beta_b^2) 2b^{\prime}} \left\{ \frac {1}{2}
\left[
\ln \left( \frac {\beta_t^2} {1 -\beta_t^2} \right) - \beta_t^2 -  \ln
(2b^{\prime}) \right] \left( 1 - \frac {1}{t^2} \right) \right.  
\nonumber \\
 &+& \left. 1 - \frac{1}{t} - \frac {\ln t}{t+1} \frac {1+2t^{\prime}}
{(1
+ t^{\prime}/2)^2} + \frac {b^{\prime 2}} {(1 + t^{\prime}/2)^2} \frac
{t-1}{2}
\right\}\ .
\label{eq:rbeb}
\end{eqnarray}
\end{widetext}
%
where $t=T/B,\ u=U/B$, $v_t$ is the speed of an electron with kinetic
energy $T$, $v_b$ is the speed of an electron with kinetic energy $B$,
$v_u$ is the speed of an electron with kinetic energy $U$, $\alpha$ is
the fine-structure constant, $a_0$ (0.529 \AA) is the Bohr radius and
\begin{equation}
\beta_t = v_t/c ,\ \ \ \ \beta_t^2 = 1 - \frac {1}{(1+t^{\prime})^2} ,\
\
\ \ t^{\prime} = T/mc^2,
\label{eq:betat}
\end{equation}
\begin{equation}
\beta_b = v_b/c ,\ \ \ \ \beta_b^2 = 1 - \frac {1}{(1+b^{\prime})^2} ,\
\
\ \ b^{\prime} = B/mc^2,
\label{eq:betab}
\end{equation}
\begin{equation}
\beta_u = v_u/c ,\ \ \ \ \beta_u^2 = 1 - \frac {1}{(1+u^{\prime})^2} ,\
\
\ \ u^{\prime} = U/mc^2 .
\label{eq:betau}
\end{equation}

The product of the ionization cross section, leading from the
S$^{(q-1)+}$ ion in the ground configuration to the S$^{q+}$ ion with
a K hole, by the statistical weight $g_i$ of the level $i$, yields the
K-shell ionization cross section $\sigma_{i}^{\mathrm{K-ion},\left(
    q-1,q\right)}$ leading from the S$^{(q-1)+}$ ion in the ground
configuration to the excited level $i$ of the S$^{q+}$ ion with a K
hole. We assumed a statistical population for each level (with a
particular $J$ value) in the final configuration.

As the electron energy distribution is not known, we used
Eq.~\eqref{eq002} and \eqref{eq003}, with cross sections computed by
us.  We assumed an electron energy distribution which is a linear
combination of a Maxwellian distribution (90\%), describing the cold
electrons at the thermodynamical temperature ($kT$=\unit{1}{keV}), and
a non-Maxwellian distribution (10\%), describing the hot electrons
($kT$=\unit{20}{keV}) heated by electron cyclotron
resonance~\cite{438}.  Details of the calculation will be given
elsewhere.  The ion densities provided by Douysset \textit{et al.}
~\cite{Douysset1} for the corresponding Ar ions were used as starting
values. We note that the initial level of the He-like M1 transition
originates from the $1s^2$ He-like ions ground configuration by excitation
and from the $1s^2 2s$ Li-like ions ground configuration by single
ionization. The initial levels of the two E1 Li-like
transitions also originate from the same configuration by excitation.

\begin{figure}
\centering
\includegraphics[clip=true,width=12cm]{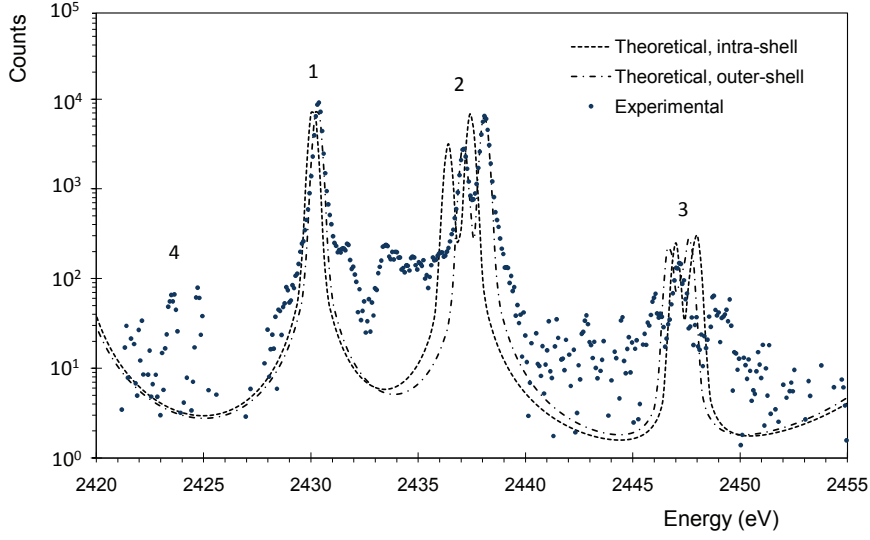}
\caption{Theoretical spectra for K excitation and single K ionization
  processes obtained with intra-shell (- - -) and outer-shell
  correlation (- $\cdot$ -) plotted against experimental data
  ({\scriptsize $\bullet$}). The peak 1 refers to the M1 1s 2s
  $^3$S$_1$ $\to$ 1s$^2$ line, the peaks 2 and 3 have
  contributions from the
  1s 2s 2p $^2$P$_{1/2,3/2}$ $\to$ 1s$^2$ 2s and the 1s
  2p $^3$P$_1$ $\to$ 1s$^2$ lines, and the peak 4 refers to the
  1s 2s 2p$^2$ $^3$S$_1$ $\to$ 1s$^2$ 2s 2p $^3$P$_{0,1,2}$
  lines.}
\label{fig:monoion}
\end{figure}

Although the theoretical spectrum thus obtained spans the
2378--\unit{2460}{eV} energy range, it is shown in
Fig.~\ref{fig:monoion} in the
2420--\unit{2455}{eV} energy range for comparison with experimental
data (dash-dotted line.) The calculated spectrum is normalized to the \unit{2438}{eV}
peak intensity.  As Fig.~\ref{fig:monoion} shows, the theoretical
spectrum resulting from excited states obtained from the ion ground
states by excitation and single K-ionization processes only is enough
to account for the major features in the experimental spectrum. In
Tab.~\ref{tab:IonDens} we list the ion densities used in the
theoretical spectrum, obtained through an iterative adjustment to the
experimental spectrum.

From the comparison with experiment we note that, although the Li-like
line intensities are well reproduced, the M1 He-like line theoretical
intensity falls below the experimental intensity. At the same time,
some minor features remain unexplained. So, we looked for processes
that could lead to other excited states suceptible of explaining the
discrepancies between our calculated values and the experimental data.

\begin{figure}[htb]
\centering
\includegraphics[clip=true,width=13cm]{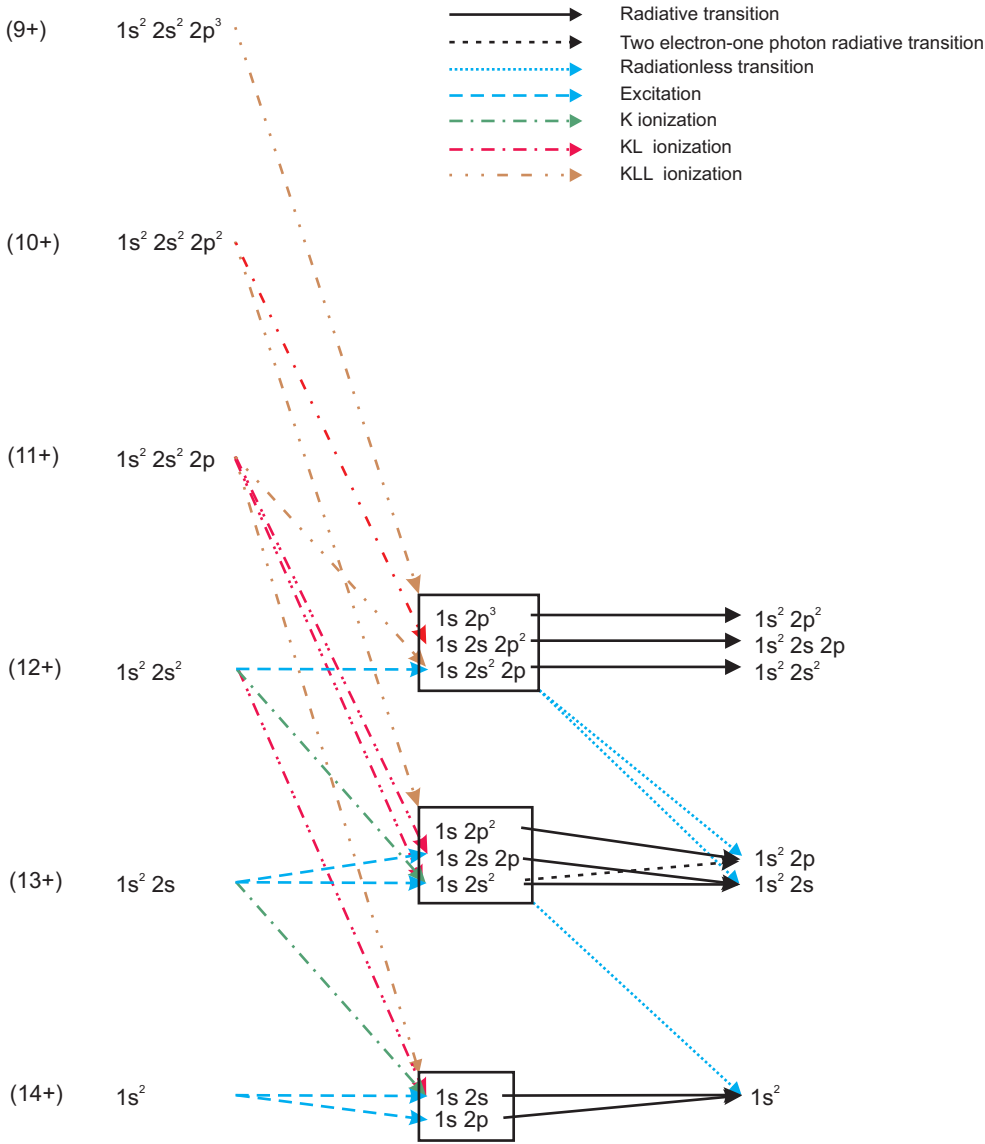}
\caption{Excitation, ionization and decay processes considered in this
  work: K excitation (- - -); single K ionization (-- $\cdot$ --);
  double KL ionization (-- $\cdot$ $\cdot$ --); triple KLL
  ionization(-- $\cdot$ $\cdot$ $\cdot$ --); radiative decay (---),
  and radiationless decay($\cdot$ $\cdot$ $\cdot$).}
\label{fig:transitions}
\end{figure}

We found out that contributions from the following configurations had
also to be included: S$^{13+}$ $1s 2p^{2}$, and S$^{12+}$ $1s 2s
2p^{2}$, and $1s 2p^{3}$.  We took into account all excitation,
single-ionization, double KL-ionization and triple KLL-ionization
processes from the ground state of S$^{9+}$ to S$^{14+}$ ions, leading
to these configurations. A diagram of all excitation, ionization and
decay processes considered in this work is shown in
Fig.~\ref{fig:transitions}.

As an example, the S$^{12+}$ $1s2s^22p\,^1P_{1}$ excited
configuration can be obtained through $1s\to2p$ excitation
from the S$^{12+}$ 1s$^{2}$\,2s$^{2}$ configuration, through K
ionization from the S$^{11+}$ $1s^{2}2s^{2}2p$ configuration,
through KL double-ionization from the S$^{10+}$ $1s^22s^22p^2$
configuration, and through KLL triple-ionization from the S$^{9+}$
$1s^22s^22p^3$ configuration.

In what concerns the calculation of the double KL and triple KLL
ionization cross-sections, we used the semi-empirical formula
developed by Shevelko and Tawara \cite{Schevelko95}, with the fitting
parameters proposed by B{\'e}lenger \textit{et al.}
\cite{Belenger97}, and the ionization energies calculated in this work
(see Tab.~\ref{tab:Ion}).

Taking in account all the referred processes, Eq.~\eqref{eq002} now reads
\begin{widetext}
\begin{eqnarray}
\label{eq004}
I_{ij}^{q} &=& 
N_{e}\hbar\omega\frac{A_{ij}^{q}}{A_{i}^{q}} 
\left[  
N_{0}^{q}\left\langle v\sigma_{i}^{\textrm{\tiny K-exc,}q}\right\rangle
+ N_{0}^{q-1}\left\langle v\sigma_{i}^{\textrm{\tiny
      K-ion,}(q-1,q)}\right\rangle \right.  \nonumber  \\ 
&+& \left.   N_{0}^{q-2}\left\langle v\sigma_{i}^{\textrm{\tiny
        KL-ion,}(q-2,q)} \right\rangle 
+ N_{0}^{q-3}\left\langle v\sigma_{i}^{\textrm{\tiny KLL-ion,}(q-3,q)} \right\rangle
 \right] . \nonumber  \\
\end{eqnarray}
\end{widetext}
In this equation, $\sigma_{i}^{\textrm{\tiny KL-ion,}(q-2,q)}$  and
$\sigma_{i}^{\textrm{\tiny KLL-ion,}(q-3,q)}$ are the double and
triple ionization cross sections respectively.  

%
%
\section{Results and discussion}
\label{sec:Res}
%

\begin{figure}[htb]
  \centering
\includegraphics[clip=true,width=13cm]{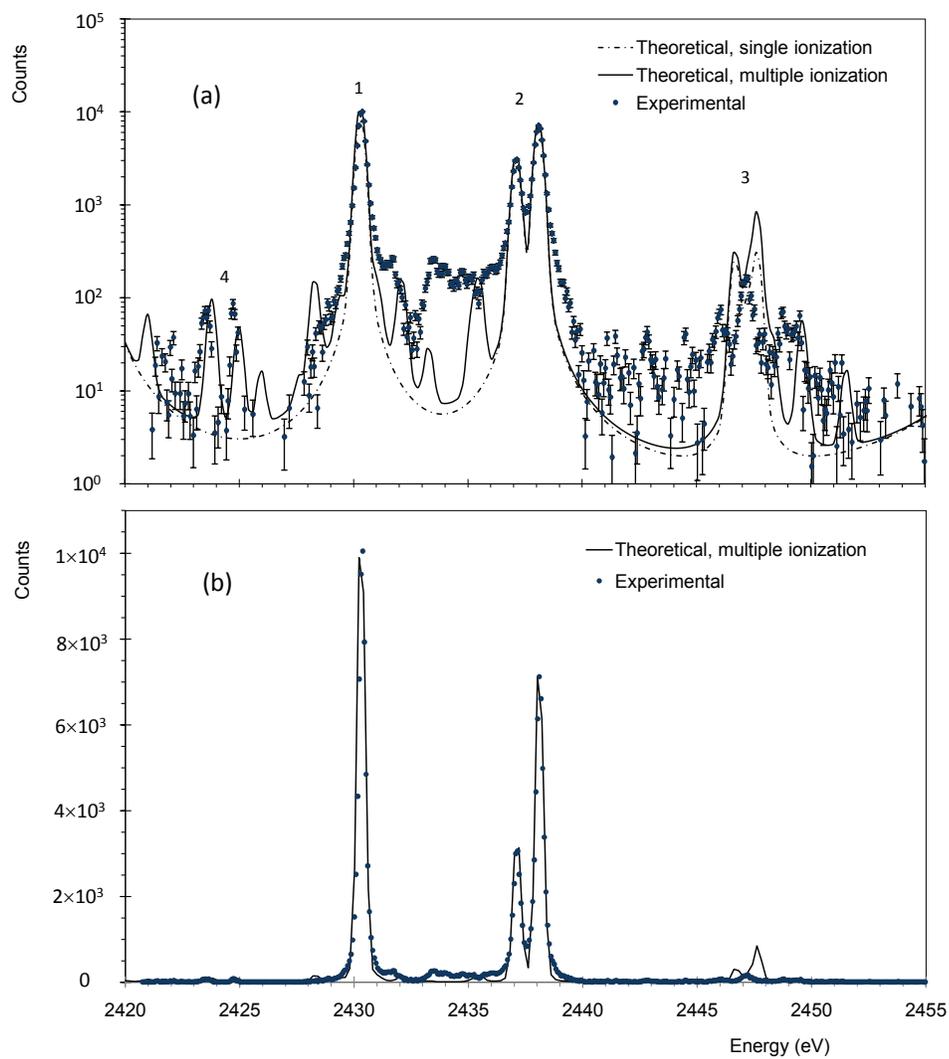}
\caption{Theoretical spectra obtained with single ionization (- $\cdot$ -)
  and multiple ionization (---) processes plotted against experimental
  data ({\scriptsize $\bullet$}) in log scale (a) and linear sclae
  (b). For the legend of the peaks, see Fig.~\ref{fig:monoion}
  caption.}
%
\label{Fig:exp_theo}
\end{figure}

The theoretical spectrum, including all lines resulting from the decay
of the excited states shown in Fig.~\ref{fig:transitions}, obtained
using the methods discussed above and assuming, for each line, a
linear combination of a gaussian and a lorentzian distributions, designed to approximate a Voigt
profile with \unit{0.3}{eV} width, is presented in Fig.~\ref{Fig:exp_theo} in a semilogarithmic
scale(a), and in a linear scale(b). 
In Fig.~\ref{Fig:exp_theo}(a) is shown, as a dash-dotted line, the
theoretical spectrum resulting from excited states obtained from the
ion ground states by excitation and single K-ionization processes
only. This is enough to account for the major features in the
experimental spectrum, clearly seen in Fig.~\ref{Fig:exp_theo}(b).
Therefore we looked for other processes that could lead to new excited
states also from the ground states. Fig.~\ref{Fig:exp_theo}(a) shows
that inclusion of double-KL and triple-KLL ionization allows for a
much better theoretical interpretation of these features.

If only single ionization and excitation processes are
considered, the $1s^2 2s$ and $1s^2$  ground configurations are the only
ones that contribute to the experimental sulphur spectrum in the
2420--\unit{2455}{eV} energy range, the first one through single ionization,
  and the second one through monopole excitation.


In this case, the \unit{2446.68}{eV} and \unit{2448.28}{eV} lines,
resulting from the decay of the $1s 2p\,^3P_{1,2}$ levels are not
significant, as these levels are poorly fed by single excitation from
the $1s^2$ ground configuration. Also based in the single excitation
and ionization processes, the calculated intensity of the  M1
line resulting from de-excitation of the $1s 2s \,^3S_1$ level is
lower than the corresponding experimental line intensity, after
adjusting the calculated spectrum to the \unit{2438}{eV} line intensity.

However, the $1s 2s$ configuration may also be obtained through
double- and triple-ionization processes, from the $1s^2 2s^2$ and
$1s^2 2s^2 2p$ configuration states, respectively. Including the
contribution of these processes in the calculated spectra allows for a
much better fit to the experimental spectra.

The intensity ratio between the M1 $1s 2s\,^3S_1\to 1s^2\,^1S_0$ and the
$1s 2s 2p\,^2P_{3/2}\to 1s^2 2s\,^2S_{1/2}$ lines (denoted 1 and 2 in
Fig.~\ref{fig:monoion}) is given by
\begin{widetext}
\begin{equation}
\frac{I_1}{I_2} = \frac{\left[
[N^{14+}\sigma^{\textrm{\tiny{exc}}}_{\textrm{\tiny{K$\to$
        L}}}({1s}^2 \to {1s 2s})+
    N^{13+}  \sigma^{\textrm{\tiny{ion}}}_{\textrm{\tiny{K}}}({1s}^2
    {2s} \to {1s 2s}) + 
N^{12+}
\sigma^{\textrm{\tiny{ion}}}_{\textrm{\tiny{KL}}}({1s}^2
{2s}^2 \to {1s 2s})\right]  TY_1} 
{N^{13+} \sigma^{\textrm{\tiny{exc}}}_{\textrm{\tiny{K$\to$
        L}}}({1s}^2 {2s} \to {1s 2s 2p}\,^2{P}_{3/2})  TY_2} 
\times g({1s 2s}\,^3{S}_1). 
\label{eq:ratio12}
\end{equation}
\end{widetext}
Here $N^{q+}$ are the densities of S$^{q+}$ ion in the ground
configuration and $g$ is the statistical weight of the $1s 2s\ ^3S_1$
level, and $TY_i$ is the transition yield of line $i$. 

The $1s 2s 2p^2$ ion configuration is obtained when the double KL and
triple KLL ionization processes are taken in account. The $1s 2s
2p^2\,^3S_1\to 1s^2 2s 2p\,^3P_{J}$, $J=0, 1, 2$ triplet visible at
the left in the figure in now predicted in agreement with experiment.

One should note that there are several weak features in the spectrum
in Fig.~\ref{Fig:exp_theo}, which are not reproduced.  These features,
are slightly above statistical noise. Some of them could be due to
$n=3$ or 4 satellite lines of the heliumlike and lithiumlike ions. Yet our
calculations did not produce any line that could explain the observed
features.  
A search in X-ray transition energy database
\cite{dkib2003} could not produce any line that could explain those
features.

The energy and transition yield values calculated in this work for all
lines contributing to the theoretical spectrum are listed in
Tab.~\ref{tab:ty}. 
The excitation cross sections are shown in Table ~\ref{tab:cs} for
\unit{10}{keV} impact energy, as an example.
%

%
\section{Conclusions}
\label{sec:Conc}
Ion densities in the plasma depend on the electron energy
distribution. For hot electrons this distribution is non-Maxwellian,
making difficult the calculation of the ion densities. So, for the
generation of the theoretical spectrum, we started from the ion
densities provided by Douysset \textit{et al} \cite{Douysset1}.
Individual line intensities were obtained using Eq.~\eqref{eq002} with
the cross section values computed by us assuming a linear combination
of Maxwellian distribution, describing the cold electrons, and a
non-Maxwellian distribution, describing the hot electrons, and the
estimated ion densities. After comparison with the experimental
spectrum, ion densities were then adjusted in order to match the
experimental peak intensities. In this way we were able to obtain a
reasonable agreement between theory and experiment.

In this work we have shown evidence of the contribution of several new
mechanism that contribute to the creation of excited state in an ECRIS
plasma. In particular double and triple processes must be considered
to reproduce properly line intensities.  We have also shown that MCDF
transition energies with medium size configuration space can reproduce
reasonably well most features of the spectra, including energies and
branching ratios. In some part of the spectrum there are hint of weak
lines that could not be reproduced so far, despite extensive search,
including $n=3$ satellites of heliumlike and lithiumlike lines.

%
%
\begin{acknowledgments}
%
%
We thank the Pionic Hydrogen collaboration
for providing us with the experimental spectra.
This research was supported in part by FCT project
POCTI/0303/2003(Portugal), financed by the European Community Fund
FEDER, by the French-Portuguese collaboration (PESSOA Program,
Contract n$^\mathrm{o}$ 10721NF), and by the Ac{\c c}{\~o}es
Integradas Luso-Francesas (Contract n$^\mathrm{o}$ F-11/09).
Laboratoire Kastler Brossel (LKB) is ``Unit\'e Mixte de Recherche du
CNRS, de l'ENS et de l'UPMC n$^{\circ}$ 8552''.  The LKB group
acknowledges the support of the Allianz Program of the Helmholtz
Association, contract EMMI HA-216 ``Extremes of Density and
Temperature: Cosmic Matter in the Laboratory''.
\end{acknowledgments}

%
%

%


\pagebreak 
\
%

%

\begin{table}
\caption{\label{tab:Corr} Electronic correlation effect
  on the 1s 2s $^3$S$_1$ $\to$ 1s$^2$ $^1$S$_0$ M1 line energy 
  and on the energy shifts, relative to the M1 line, of the 1s 2s
  2p $^2$P$_{1/2}$ $\to$ 1s$^2$ 2s E1(1) and   1s 2s
  2p $^2$P$_{3/2}$ $\to$ 1s$^2$ 2s E1(2) lines, compared to
  experiment. The energy values are in eV.
}
\begin{ruledtabular}
\begin{tabular}{lccc}
            & M1 & E1 (1) & E1 (2) \\ 
\hline
Intra-shell correlation (up to 2p)       &       2430.09 &       6.30    &       7.34    \\
Outer-shell correlation (up to 4f)       &       2430.29 &       6.82    &       7.79    \\
Experiment      &               &       6.76    &       7.75    \\
\end{tabular}
\end{ruledtabular}
\end{table}
%
%
%

\begin{table}
\caption{\label{tab:IonDens}Ion density values obtained
  through an iterative adjustment to the experimental spectrum.  
}
\begin{ruledtabular}
\begin{tabular}{lc}
        &  Ion densities (m$^{-3}$)   \\
\hline                                                                                  
S$^{9+} $  &   $8.0\times^{15}$   \\
S$^{10+}$  &   $7.0\times^{15}$   \\
S$^{11+}$  &   $7.0\times^{15}$   \\
S$^{12+}$  &   $9.0\times^{15}$   \\
S$^{13+}$  &   $1.4\times^{15}$   \\
S$^{14+}$  &   $7.0\times^{15}$   \\
\end{tabular}
\end{ruledtabular}
\end{table}
%
%
%

\begin{table}
\caption{\label{tab:Ion}Calculated energies for single K- (1s$^{-1}$), double KL-
  (1s$^{-1}$ 2s$^{-1}$, 1s$^{-1}$ 2p$^{-1}$) and triple KLL-
  (1s$^{-1}$ 2s$^{-2}$) ionization (in eV). 
}
\begin{ruledtabular}
\begin{tabular}{llcccc}
        &       Configuration   &       1s$^{-1}$       &       1s$^{-1}$ 2s$^{-1}$     &       1s$^{-1}$ 2p$^{-1}$     &       1s$^{-1}$ 2s$^{-2}$     \\
\hline                                                                                  
S$^{9+}$        &       1s$^2$  2s$^2$ 2p$^3$   &       2799.63 &       3332.02 &       3328.31 &       4344.02 \\
S$^{10+}$       &       1s$^2$  2s$^2$ 2p$^2$   &       2880.79 &       3476.37 &       3472.45 &       4608.55 \\
S$^{11+}$       &       1s$^2$  2s$^2$ 2p       &       2967.42 &       3626.01 &       3619.21 &       4869.29 \\
S$^{12+}$       &       1s$^2$  2s$^2$  &       3059.32 &       3787.89 &               &       5141.14 \\
S$^{13+}$       &       1s$^2$  2s      &       3135.92 &       3929.29 &               &               \\
S$^{14+}$       &       1s$^2$  &       3222.42 &               &               &               \\
\end{tabular}
\end{ruledtabular}
\end{table}
%
%
\pagebreak
 
\begin{longtable}{clclccc}
\caption{Energy and Transition yield ($TY$) values calculated in this work
  for all lines contributing to the theoretical spectrum.} \label{tab:ty}\\
[-1.0ex]
%
   \hline \hline \\
   [-2ex]
   $q$ & conf$_i$              & $LSJ_i$   & conf$_f$     & $LSJ_f$    & $E$ (eV)  & $TY$ \\
   \hline
   \\
   [-1.8ex]
\endfirsthead
%
\multicolumn{7}{c}{{\tablename} \thetable{} -- Continued} \\
  \hline \hline \\
  [-2ex]
  $q$ & conf$_i$              & $LSJ_i$   & conf$_f$     & $LSJ_f$    & $E$ (eV)  & $TY$ \\
  \hline
\endhead

  \multicolumn{7}{l}{{Continued on Next Page\ldots}} \\
  \hline \hline 
\endfoot

  \\[-1.8ex] \hline \hline
\endlastfoot

%
13+     &       1s  2s$^2$      &       $^2$S$_{1/2}$   &       1s$^2$ 2p       &       $^2$P$_{3/2}$   &       2378.45 &       1.75E-02        \\
13+     &       1s  2s$^2$      &       $^2$S$_{1/2}$   &       1s$^2$ 2p       &       $^2$P$_{1/2}$   &       2380.36 &       1.02E-02        \\
12+     &       1s  2p$^3$      &       $^5$S$_2$       &       1s$^2$  2p$^2$  &       $^3$P$_2$       &       2387.45 &       1.06E-03        \\
12+     &       1s  2p$^3$      &       $^5$S$_2$       &       1s$^2$  2p$^2$  &       $^3$P$_1$       &       2388.45 &       7.10E-04        \\
12+     &       1s  2p$^3$      &       $^3$D$_3$       &       1s$^2$  2p$^2$  &       $^1$D$_2$       &       2397.23 &       1.97E-03        \\
12+     &       1s  2p$^3$      &       $^3$D$_3$       &       1s$^2$  2p$^2$  &       $^3$P$_2$       &       2404.12 &       1.27E-01        \\
12+     &       1s  2p$^3$      &       $^1$P$_1$       &       1s$^2$  2p$^2$  &       $^1$S$_0$       &       2404.20 &       3.39E-01        \\
12+     &       1s  2p$^3$      &       $^3$D$_2$       &       1s$^2$  2p$^2$  &       $^3$P$_2$       &       2404.39 &       7.82E-02        \\
12+     &       1s  2p$^3$      &       $^3$D$_1$       &       1s$^2$  2p$^2$  &       $^3$P$_2$       &       2404.39 &       8.66E-03        \\
12+     &       1s  2s  2p$^2$  &       $^1$D$_2$       &       1s$^2$ 2s  2p   &       $^1$P$_1$       &       2405.31 &       3.93E-01        \\
12+     &       1s  2p$^3$      &       $^3$D$_2$       &       1s$^2$  2p$^2$  &       $^3$P$_1$       &       2405.40 &       3.80E-01        \\
12+     &       1s  2p$^3$      &       $^3$D$_1$       &       1s$^2$  2p$^2$  &       $^3$P$_1$       &       2405.40 &       1.29E-01        \\
12+     &       1s  2s$^2$ 2p   &       $^3$P$_1$       &       1s$^2$ 2s$^2$   &       $^1$S$_0$       &       2405.75 &       3.37E-03        \\
12+     &       1s  2s  2p$^2$  &       $^3$P$_1$       &       1s$^2$ 2s  2p   &       $^1$P$_1$       &       2406.00 &       3.82E-02        \\
12+     &       1s  2p$^3$      &       $^3$D$_1$       &       1s$^2$  2p$^2$  &       $^3$P$_0$       &       2406.04 &       2.31E-01        \\
12+     &       1s  2s  2p$^2$  &       $^3$P$_2$       &       1s$^2$ 2s  2p   &       $^1$P$_1$       &       2407.05 &       1.93E-01        \\
12+     &       1s  2p$^3$      &       $^1$D$_2$       &       1s$^2$  2p$^2$  &       $^1$D$_2$       &       2408.18 &       7.78E-01        \\
12+     &       1s  2p$^3$      &       $^3$S$_1$       &       1s$^2$  2p$^2$  &       $^3$P$_2$       &       2409.99 &       3.73E-01        \\
12+     &       1s  2p$^3$      &       $^3$P$_2$       &       1s$^2$  2p$^2$  &       $^1$D$_2$       &       2410.23 &       2.99E-01        \\
12+     &       1s  2p$^3$      &       $^3$S$_1$       &       1s$^2$  2p$^2$  &       $^3$P$_1$       &       2411.00 &       2.82E-01        \\
12+     &       1s  2s  2p$^2$  &       $^3$D$_3$       &       1s$^2$ 2s  2p   &       $^3$P$_2$       &       2411.10 &       3.28E-01        \\
12+     &       1s  2s  2p$^2$  &       $^3$D$_2$       &       1s$^2$ 2s  2p   &       $^3$P$_1$       &       2411.41 &       2.94E-01        \\
12+     &       1s  2p$^3$      &       $^3$S$_1$       &       1s$^2$  2p$^2$  &       $^3$P$_0$       &       2411.64 &       9.44E-02        \\
12+     &       1s  2s  2p$^2$  &       $^3$P$_1$       &       1s$^2$ 2s  2p   &       $^3$P$_2$       &       2412.07 &       2.75E-01        \\
12+     &       1s  2s  2p$^2$  &       $^3$D$_2$       &       1s$^2$ 2s  2p   &       $^3$P$_2$       &       2412.22 &       9.74E-03        \\
12+     &       1s  2s  2p$^2$  &       $^3$P$_0$       &       1s$^2$ 2s  2p   &       $^3$P$_1$       &       2412.29 &       5.29E-01        \\
12+     &       1s  2s  2p$^2$  &       $^3$D$_1$       &       1s$^2$ 2s  2p   &       $^3$P$_2$       &       2412.39 &       1.39E-01        \\
12+     &       1s  2s  2p$^2$  &       $^3$P$_2$       &       1s$^2$ 2s  2p   &       $^3$P$_2$       &       2413.04 &       8.62E-01        \\
12+     &       1s  2s  2p$^2$  &       $^3$P$_1$       &       1s$^2$ 2s  2p   &       $^3$P$_1$       &       2413.27 &       3.83E-01        \\
12+     &       1s  2s  2p$^2$  &       $^3$D$_1$       &       1s$^2$ 2s  2p   &       $^3$P$_1$       &       2413.36 &       2.70E-01        \\
12+     &       1s  2s  2p$^2$  &       $^3$P$_2$       &       1s$^2$ 2s  2p   &       $^3$P$_1$       &       2414.23 &       2.76E-02        \\
12+     &       1s  2s  2p$^2$  &       $^3$P$_1$       &       1s$^2$ 2s  2p   &       $^3$P$_0$       &       2414.24 &       3.69E-02        \\
12+     &       1s  2p$^3$      &       $^1$D$_2$       &       1s$^2$  2p$^2$  &       $^3$P$_2$       &       2415.07 &       1.75E-01        \\
13+     &       1s  2s  2p      &       $^4$P$_{1/2}$   &       1s$^2$ 2s       &       $^2$S$_{1/2}$   &       2415.18 &       5.28E-01        \\
13+     &       1s  2s  2p      &       $^4$P$_{3/2}$   &       1s$^2$ 2s       &       $^2$S$_{1/2}$   &       2415.69 &       2.14E-02        \\
%
%
12+     &       1s  2s  2p$^2$  &       $^1$P$_1$       &       1s$^2$ 2s  2p   &       $^1$P$_1$       &       2416.02 &       9.80E-01        \\
12+     &       1s  2p$^3$      &       $^1$D$_2$       &       1s$^2$  2p$^2$  &       $^3$P$_1$       &       2416.07 &       1.66E-02        \\
12+     &       1s  2p$^3$      &       $^3$P$_1$       &       1s$^2$  2p$^2$  &       $^3$P$_2$       &       2416.49 &       5.44E-01        \\
13+     &       1s  2s  2p      &       $^4$P$_{5/2}$   &       1s$^2$ 2s       &       $^2$S$_{1/2}$   &       2416.93 &       1.21E-01        \\
13+     &       1s  2p$^2$      &       $^4$P$_{5/2}$   &       1s$^2$ 2p       &       $^2$P$_{3/2}$   &       2417.01 &       9.90E-01        \\
12+     &       1s  2p$^3$      &       $^3$P$_2$       &       1s$^2$  2p$^2$  &       $^3$P$_2$       &       2417.13 &       1.06E-01        \\
13+     &       1s  2p$^2$      &       $^4$P$_{1/2}$   &       1s$^2$ 2p       &       $^2$P$_{1/2}$   &       2417.16 &       9.29E-03        \\
12+     &       1s  2p$^3$      &       $^3$P$_1$       &       1s$^2$  2p$^2$  &       $^3$P$_1$       &       2417.49 &       1.26E-01        \\
12+     &       1s  2p$^3$      &       $^3$P$_0$       &       1s$^2$  2p$^2$  &       $^3$P$_1$       &       2417.56 &       1.00E+00        \\
12+     &       1s  2p$^3$      &       $^3$P$_2$       &       1s$^2$  2p$^2$  &       $^3$P$_1$       &       2418.13 &       3.24E-02        \\
12+     &       1s  2p$^3$      &       $^3$P$_1$       &       1s$^2$  2p$^2$  &       $^3$P$_0$       &       2418.14 &       1.46E-01        \\
12+     &       1s  2s  2p$^2$  &       $^1$S$_0$       &       1s$^2$ 2s  2p   &       $^1$P$_1$       &       2418.49 &       4.64E-01        \\
12+     &       1s  2s$^2$ 2p   &       $^1$P$_1$       &       1s$^2$ 2s$^2$   &       $^1$S$_0$       &       2418.58 &       3.41E-01        \\
12+     &       1s  2p$^3$      &       $^1$P$_1$       &       1s$^2$  2p$^2$  &       $^1$D$_2$       &       2420.98 &       3.85E-01        \\
12+     &       1s  2s  2p$^2$  &       $^3$S$_1$       &       1s$^2$ 2s  2p   &       $^3$P$_2$       &       2423.78 &       3.72E-01        \\
12+     &       1s  2s  2p$^2$  &       $^3$S$_1$       &       1s$^2$ 2s  2p   &       $^3$P$_1$       &       2424.97 &       1.80E-01        \\
12+     &       1s  2s  2p$^2$  &       $^3$S$_1$       &       1s$^2$ 2s  2p   &       $^3$P$_0$       &       2425.94 &       5.06E-02        \\
12+     &       1s  2s  2p$^2$  &       $^1$D$_2$       &       1s$^2$ 2s  2p   &       $^3$P$_2$       &       2427.61 &       1.50E-02        \\
12+     &       1s  2s  2p$^2$  &       $^3$P$_1$       &       1s$^2$ 2s  2p   &       $^3$P$_2$       &       2428.29 &       6.35E-01        \\
12+     &       1s  2s  2p$^2$  &       $^3$P$_0$       &       1s$^2$ 2s  2p   &       $^3$P$_1$       &       2428.68 &       8.25E-02        \\
12+     &       1s  2s  2p$^2$  &       $^3$P$_2$       &       1s$^2$ 2s  2p   &       $^3$P$_2$       &       2429.35 &       1.19E-01        \\
12+     &       1s  2s  2p$^2$  &       $^3$P$_1$       &       1s$^2$ 2s  2p   &       $^3$P$_1$       &       2429.48 &       2.85E-02        \\
14+     &       1s  2s  &       $^3$S$_1$       &       1s$^2$  &       $^1$S$_0$       &       2430.29 &       1.00E+00        \\
12+     &       1s  2s  2p$^2$  &       $^3$P$_1$       &       1s$^2$ 2s  2p   &       $^3$P$_0$       &       2430.45 &       6.82E-02        \\
12+     &       1s  2s  2p$^2$  &       $^3$P$_2$       &       1s$^2$ 2s  2p   &       $^3$P$_1$       &       2430.54 &       2.88E-02        \\
13+     &       1s  2p$^2$      &       $^2$D$_{5/2}$   &       1s$^2$ 2p       &       $^2$P$_{3/2}$   &       2431.02 &       3.53E-01        \\
13+     &       1s  2p$^2$      &       $^2$D$_{3/2}$   &       1s$^2$ 2p       &       $^2$P$_{1/2}$   &       2432.09 &       9.94E-01        \\
13+     &       1s  2p$^2$      &       $^2$P$_{1/2}$   &       1s$^2$ 2p       &       $^2$P$_{3/2}$   &       2433.28 &       2.80E-01        \\
13+     &       1s  2p$^2$      &       $^2$P$_{1/2}$   &       1s$^2$ 2p       &       $^2$P$_{1/2}$   &       2435.19 &       6.57E-01        \\
13+     &       1s  2p$^2$      &       $^2$P$_{3/2}$   &       1s$^2$ 2p       &       $^2$P$_{3/2}$   &       2435.49 &       9.22E-01        \\
13+     &       1s  2s  2p      &       $^2$P$_{1/2}$   &       1s$^2$ 2s       &       $^2$S$_{1/2}$   &       2437.11 &       9.39E-01        \\
13+     &       1s  2p$^2$      &       $^2$P$_{3/2}$   &       1s$^2$ 2p       &       $^2$P$_{1/2}$   &       2437.40 &       7.83E-02        \\
13+     &       1s  2s  2p      &       $^2$P$_{3/2}$   &       1s$^2$ 2s       &       $^2$S$_{1/2}$   &       2438.08 &       1.00E+00        \\
14+     &       1s  2p  &       $^3$P$_1$       &       1s$^2$  &       $^1$S$_0$       &       2446.68 &       1.00E+00        \\
13+     &       1s  2s  2p      &       $^2$P$_{1/2}$   &       1s$^2$ 2s       &       $^2$S$_{1/2}$   &       2447.24 &       3.24E-01        \\
13+     &       1s  2s  2p      &       $^2$P$_{3/2}$   &       1s$^2$ 2s       &       $^2$S$_{1/2}$   &       2447.64 &       9.06E-01        \\
14+     &       1s  2p  &        $^3$P$_2$      &       1s$^2$  &       $^1$S$_0$       &       2448.28 &       3.10E-01        \\
13+     &       1s  2p$^2$      &       $^2$S$_{1/2}$   &       1s$^2$ 2p       &       $^2$P$_{3/2}$   &       2449.62 &       6.60E-01        \\
13+     &       1s  2p$^2$      &       $^2$S$_{1/2}$   &       1s$^2$ 2p       &       $^2$P$_{1/2}$   &       2451.53 &       1.89E-01        \\
14+     &       1s  2p  &       $^1$P$_1$       &       1s$^2$  &       $^1$S$_0$       &       2460.15 &       1.00E+00        \\
\end{longtable}
%
%

\pagebreak
\begin{table}[tbp]
\caption{\label{tab:cs}Sulphur ions excitation cross sections, for
        10 keV electron impact energy.  
}
\begin{ruledtabular}
\begin{tabular}{cccccc}
$q$     &       initial conf.  &   & final conf. &    & $\sigma$ (m$^2$)        \\
\hline                                                                                  
$12^+$  & 1s$^2$ 2s$^2$   &       $^1$S$_0$       &       1s 2s$^2$ 2p    &       $^3$P$_1$       &       $2.05 \times 10^{-27}$  \\
        &                 &               &               &       $^1$P$_1$       &       $3.05 \times 10^{-25}$  \\
\hline                                                                                  
$13^+$  & 1s$^2$ 2s       &       $^2$S$_{1/2}$   &       1s 2s$^2$       &       $^2$S$_{1/2}$   &       $1.96 \times 10^{-26}$  \\
        &                 &                       &       1s 2s 2p        &       $^4$P$_{1/2}$   &       $6.68 \times 10^{-29}$  \\
        &                 &               &               &       $^2$P$_{1/2}^ {(2)}$    &       $1.62 \times 10^{-25}$  \\
        &                 &               &               &       $^2$P$_{1/2}^{(1)}$     &       $1.64 \times 10^{-26}$  \\
        &                 &               &               &       $^4$P$_{3/2}$   &       $3.68 \times 10^{-28}$  \\
        &                 &               &               &       $^2$P$_{3/2}^{(1)}$     &       $3.45 \times 10^{-25}$  \\
        &                 &               &               &       $^2$P$_{3/2}^{(2)}$     &       $1.06 \times 10^{-26}$  \\
\hline                                                                                  
$14^+$  &       1s$^2$    &       $^1$S$_0$       &       1s 2p   &       $^3$P$_1$       &       $2.29\times 10^{-27}$   \\
        &                 &               &               &       $^1$P$_1$       &       $3.21 \times 10^{-25}$  \\
        &                 &               &               &       $^3$S$_1$       &  $3.72 \times 10^{-26}$       \\
\end{tabular}
\end{ruledtabular}
\end{table}
%
%
%
%
%
%
%

\pagebreak 
\
\cleardoublepage

%
%
%
%

%
%

\end{document}